\documentclass[prb,twocolumn,groupedaddress]{revtex4}

\usepackage[dvips]{color}
\usepackage{graphicx}
\usepackage {amssymb}

\begin{document}

\title{Scanning Tunneling Spectroscopy of Suspended Single-Wall Carbon Nanotubes}

\author{B.J. LeRoy}
\email[Electronic mail: ]{leroy@mb.tn.tudelft.nl}
\author{S.G. Lemay}
\author{J. Kong}
\author{C. Dekker}
\affiliation{Department of Nanoscience, Delft University of
Technology, Lorentzweg 1, 2628 CJ Delft, The Netherlands}

\date{\today}

\begin{abstract}
We have performed low-temperature STM measurements on single-wall
carbon nanotubes that are freely suspended over a trench. The
nanotubes were grown by CVD on a Pt substrate with predefined
trenches etched into it. Atomic resolution was obtained on the
freestanding portions of the nanotubes. Spatially resolved
spectroscopy on the suspended portion of both metallic and
semiconducting nanotubes was also achieved, showing a
Coulomb-staircase behavior superimposed on the local density of
states. The spacing of the Coulomb blockade peaks changed with tip
position reflecting a changing tip-tube capacitance.
\end{abstract}

\maketitle

Transport measurements on single-wall carbon nanotubes (SWCNTs)
show Coulomb blockade \cite{Bockrath97, Tans} and Luttinger liquid
behavior.\cite{Bockrath99, Yao}  While it would be desirable to
use the high spatial resolution of scanning tunneling microscopy
(STM) to study these effects, they are obscured by the close
proximity of a conducting substrate.  Suspending SWCNTs can
circumvent this limitation.  Several ways of suspending SWCNTs for
transport measurements have been reported\cite{Tombler, Nygard}
but none of these is compatible with STM.

In this letter, we demonstrate that it is possible to obtain
atomic-resolution STM images and perform local spectroscopy
measurements on suspended carbon nanotubes.  The tubes are grown
across trenches on a metallic substrate to allow STM imaging.
Spatially resolved spectroscopy on metallic and semiconducting
nanotubes shows a Coulomb staircase due to the addition of single
electrons. The spacing between steps in the staircase corresponds
to the energy necessary to add an electron.  By measuring this
spacing, we determine the capacitance between the tip and
nanotube, which is found to depend on the distance of the tip from
the edge of the trench.

\begin{figure}
\begin{center}
\includegraphics[width=3.35in]{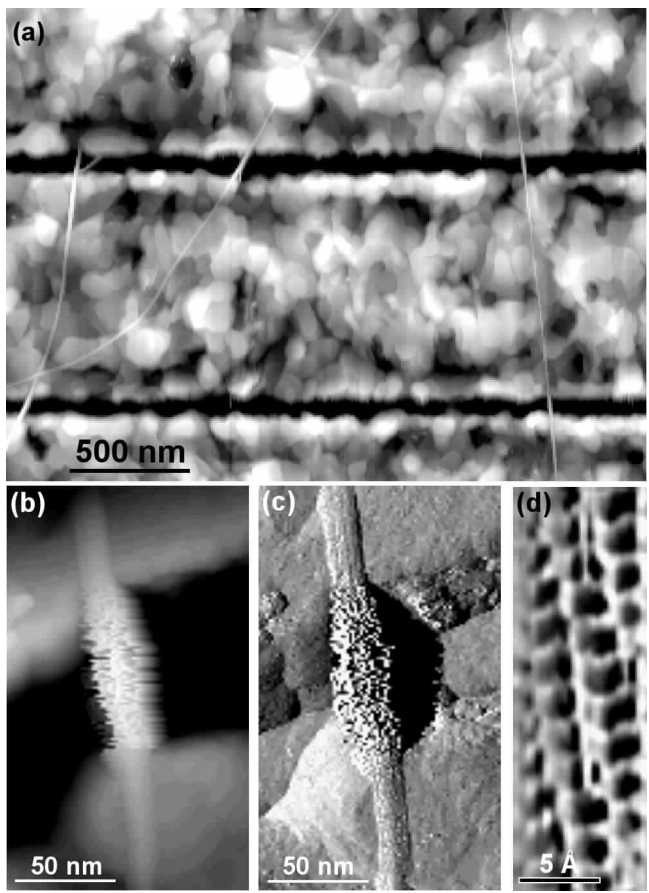}%
\caption{(a) AFM image of the sample structure showing nanotubes
crossing 100 nm wide trenches.  (b) STM topography of a nanotube
crossing a trench showing the freely suspended portion of the
tube.  (c) STM current image showing current spikes on the
suspended portion of the nanotube. (d) High-resolution topography
on a suspended portion of the nanotube showing the atomic
structure. The vertical axis is parallel to both the tube axis and
the scan direction. All three STM images were taken with a sample
voltage of -0.5 V and a feedback current of 300 pA.} \label{fig:1}
\end{center}
\end{figure}

Figure \ref{fig:1}(a) shows an Atomic Force Microscopy (AFM) image
of the structure with suspended carbon nanotubes.  100 nm wide
trenches were dry-etched in SiO$_2$ to a depth of 200 nm. The
spacing between trenches was 1 $\mu$m. After the etching, a 100 nm
thick film of Pt was deposited onto the entire sample by
evaporation to create a conducting substrate. This was followed by
deposition of 5 $\mu$m square areas of Fe:Mo catalyst. Nanotubes
were grown from the catalyst by CVD at 800 $^{\circ}$C for ten
minutes.\cite{Kong}  The tubes grow from the catalyst in random
directions; some of them cross over the trenches and are therefore
suspended for a distance of approximately 100 nm. The nanotubes
can be seen as narrow straight lines running over the trenches in
several locations of Fig. \ref{fig:1}(a).

The samples were measured using a UHV low-temperature STM with a
base temperature of 4.7 K (Omicron LT-STM).  STM tips were
mechanically cut from Pt-Ir wire.  The tips are sufficiently blunt
($\approx$ 60 nm radius of curvature) that the apex never reaches
the bottom of the trench. Figure \ref{fig:1}(b) is a STM
topography image in constant current mode of a nanotube crossing a
trench. The dark horizontal area is the trench, which the nanotube
crosses. Figure \ref{fig:1}(c) shows the corresponding current
image.  The nanotube appears slightly higher in the region over
the trench due to attractive forces between the tip and nanotube.
This attraction also leads to sharp spikes in the current because
of the movement of the nanotube. The dark area (no current) to the
right of the nanotube is where the tip loses contact with the tube
and where it has not yet reached the trench.

Figure \ref{fig:1}(d) is an image showing atomic resolution on the
free-standing portion of the nanotube.  The height of the atoms
measured in constant current mode was approximately 4 \AA. This is
about five times larger than on supported nanotubes. The large
apparent height of the atoms may be due to the ability of the
nanotube to move due to the forces acting on it from the tip.
These images demonstrate the ability to image and obtain atomic
resolution on free-standing nanotubes.

\begin{figure}
\begin{center}
\includegraphics[width=3.0in]{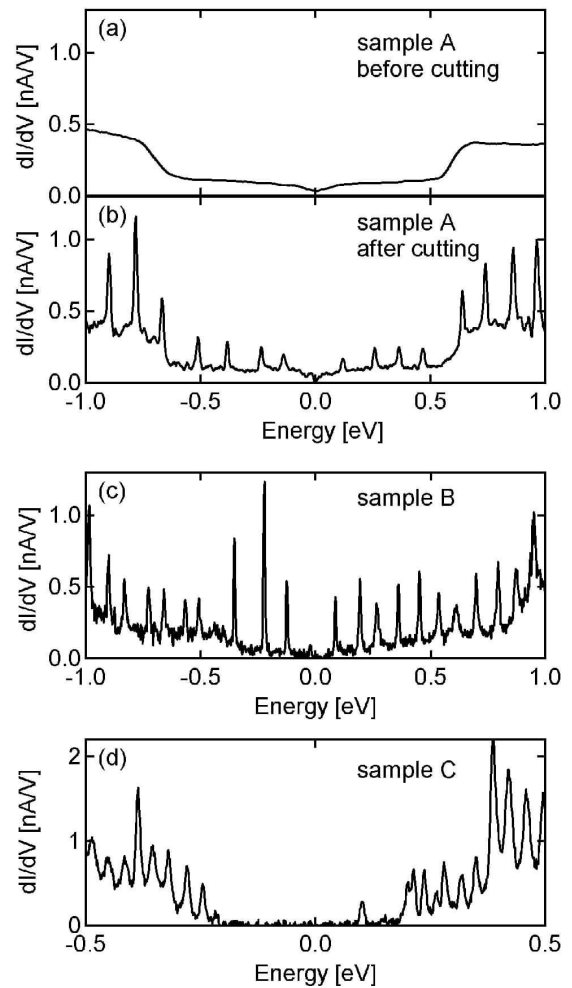}%
\caption{Spectroscopy on suspended metallic tube (a) before and
(b) after cutting.  The sharp spikes from the Coulomb Staircase
appeared after cutting.  (c) Spectroscopy on a different suspended
metallic nanotube.  (d) Spectroscopy on a semiconducting nanotube
showing the gap as well as the Coulomb staircase.} \label{fig:2}
\end{center}
\end{figure}

Figure \ref{fig:2}(a) shows spectroscopy on a suspended nanotube.
The spectroscopy curves were obtained using lock-in detection (867
Hz). We observe a constant density of states at low energy and the
first van Hove singularities are visible at higher energies,
characteristic of a metallic nanotube.\cite{Wildoer, Odom}  The
nanotube was then cut to a shorter length by applying voltage
pulses to it on both sides of the trench, reducing its length to
140 nm.\cite{Venema97} Figure \ref{fig:2}(b) plots the
differential conductance after cutting. Sharp spikes have appeared
superimposed on the background density of states. The spike
pattern can be attributed to Coulomb blockade, where the $dI/dV$
peaks are caused by the addition of single electrons to the SWCNT.

The Coulomb staircase is characterized by the resistances and
capacitances of the two tunnel barriers.  These determine the
slope of the $I-V$ curve and also the spacing between peaks in
$dI/dV$.  The resistances and capacitances in the system can be
determined from the spectroscopy measurements.\cite{Grabert,
Hanna}  The tunneling gap between the tip and the SWCNT ensures
that $R_{tip} \gg R_{sub}$, where $R_{tip}$ is the resistance
between the tip and the SWCNT and $R_{sub}$ is between the SWCNT
and substrate. Therefore, $R_{tip}$ approximately equals the total
resistance as determined by the tunneling setpoint.  The
capacitance between the tip and the SWCNT, $C_{tip}$ is determined
from the spacing $\Delta V$ between peaks in the Coulomb
staircase,
\begin{equation}
\label{cap} \Delta V \approx e / C_{tip}.
\end{equation}
Here we have neglected the level spacing in the SWCNT, which
causes a small variation in the values of $\Delta V$ as discussed
below. The capacitance between the SWCNT and the substrate,
$C_{sub}$ is determined from the slope between steps in the $I-V$
curves. For the nanotube of Fig. \ref{fig:2} (b), $C_{tip}
\approx$ 1 aF and $C_{sub} \approx$ 10 aF.  The value for
$C_{tip}$ is in agreement with numerical simulations, where the
tip is modelled as a 60 nm radius sphere. A rough estimate for the
capacitance of a cylinder lying above a metal plane gives a value
of 0.06 aF/nm, or $C_{sub} \approx$ 2.4 aF for our nanotube,
consistent with the observed order of magnitude. We have observed
that reducing the tunnel current increases the peak spacing.  This
is consistent with the increased distance between tip and SWCNT
decreasing $C_{tip}$. We have also observed that the height of the
spikes in $dI/dV$ is proportional to the density of states; the
peaks increase in height at energies greater than the first van
Hove singularities. This is also consistent with the Coulomb
staircase model.

Figure \ref{fig:2}(c) shows spectroscopy on another metallic
nanotube.  In this case a Coulomb staircase was observed without
cutting the nanotube. This is the behavior that is seen more often
in our nanotubes; we have observed the Coulomb staircase on all 12
SWCNTs that we have measured, with 9 showing it without being cut.
It is attributed to the presence of local defects (tunnel
barriers) induced by the edges of the trench. These may be caused
by the nanotube bending at the edge of the trench, therefore
creating a tunnel barrier. When the tunnel barriers are present,
the relevant length for the nanotube device is approximately the
distance between the two metal contacts. Once again, we can
determine the capacitances from the spectroscopy data.  The
measured capacitances were $C_{tip} \approx$ 2 aF and $C_{sub}
\approx$ 1 aF. The low value for $C_{sub}$ further supports the
conclusion that there is a tunnel barrier near the edge of the
trench, isolating the portion of the nanotube that is over the
trench from the part that is on the Pt.

Figure \ref{fig:2}(d) shows spectroscopy on a suspended
semiconducting nanotube.  At low energy, there is a gap while at
higher energy there is a finite density of states.  Once the
energy is sufficient to overcome the gap, the spikes from the
Coulomb staircase appear superimposed on the local density of
states. This shows that the Coulomb staircase is not particular to
the metallic nanotubes but rather is a generic feature of
suspended nanotubes.

\begin{figure}
\begin{center}
\includegraphics[width=3.35in]{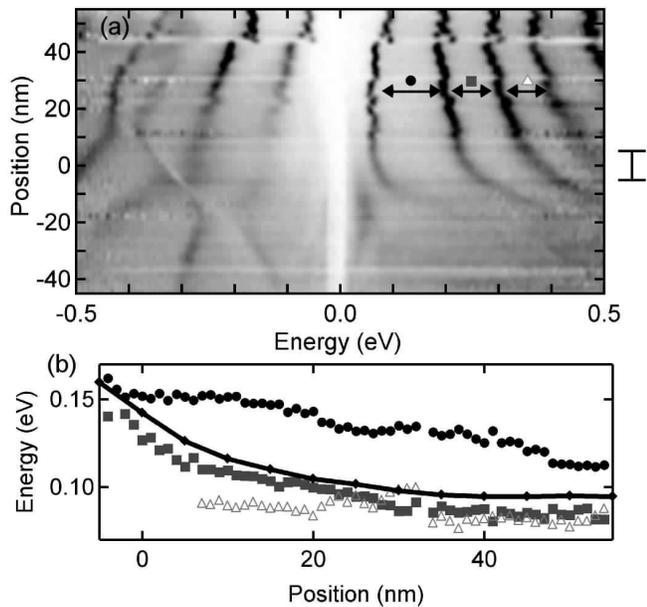}%
\caption{(a) Tunneling spectroscopy as a function of tip position
for a suspended metallic nanotube.  The locations of the peaks
changes with position, indicating that the tip-tube capacitance is
changing.  The estimated position of the edge of the trench is
marked on the right. (b) Peak spacing as a function of position
for the first four peaks on the positive-bias side of (a).
\Large$\bullet$ \small represents the spacing between peaks 1 and
2, $\blacksquare$ is for peaks 2 and 3 and the $\vartriangle$ is
for peaks 3 and 4.  The solid line is a numerical simulation of
the capacitance between the tip and SWCNT.} \label{fig:3}
\end{center}
\end{figure}

We have also performed spatially resolved spectroscopy on
nanotubes going over a trench.  Figure ~\ref{fig:3}(a) plots the
local density of states as a function of position in a metallic
nanotube.  Our best estimate for the position of the edge of the
trench from topography measurements is indicated by the bracket on
the right of the image. The dark lines correspond to high
conductance and the addition of electrons. The spacing between
peaks in the Coulomb staircase, $\Delta V$, is determined by the
capacitance between the tip and the nanotube (Eq. \ref{cap}).  The
increased spacing of the Coulomb peaks as the tip approaches the
edge of the trench implies that the capacitance $C_{tip}$ is
decreasing. This is due to screening of the field from the tip by
the metal of the substrate. The spacing between successive peaks
thus gives a local measurement of the capacitance between the tip
and nanotube.  We observed no change in $C_{sub}$ along the length
of the nanotube.

Figure \ref{fig:3}(b) shows the energy spacing between the Coulomb
peaks as a function of position along the nanotube.  We have
performed a numerical simulation of the capacitance between the
tip and nanotube.  The calculated capacitances have been converted
to a peak spacing using Eq. \ref{cap}.  The solid line plots the
result showing qualitative agreement with the measured peak
spacing. The simulations predict a 30\% decrease in the
capacitance when the tip is at the edge of the trench compared to
the center.

We attribute the different spacings between peaks to the level
spacing in the nanotube.  The average spacing between levels in a
nanotube is given by $\Delta E = hv_F/4L$ where $v_F$ is the Fermi
velocity and $h$ is Planck's constant.  For a 100 nm nanotube,
$\Delta E$ is 9 meV.  The measured energy difference between peaks
must be scaled by the fraction of the voltage that drops across
the tip-nanotube junction, $C_{sub} / (C_{sub}+C_{tip})$. This is
a factor of 1/3 for the nanotube of Fig. \ref{fig:3}(b), giving a
$\Delta E$ of 12 meV in good agreement with the expected $\Delta
E$ for a 100 nm nanotube.

We have demonstrated that it is possible to obtain atomic
resolution and perform spectroscopy measurements on suspended
individual single-walled carbon nanotubes.  The suspended portions
of the nanotubes showed a Coulomb staircase, which is not observed
in conventional STM measurements due to the large $C_{sub}$.  The
Coulomb staircase allows a local determination of the capacitance
between the tip and nanotube. The ability to fabricate and image
devices with freely suspended nanotubes will allow simultaneous
transport and scanning tunneling spectroscopy measurements.

The authors would like to thank NWO Pionier and FOM for funding,
and Abdou Hassanien who was involved in earlier stages of this
project.

\bibliography{LeRoyAPL}







\end{document}